\documentclass[10pt,english,pre,nofootinbib,reprint, groupedaddress,showpacs,showkeys,preprintnumbers,floatfix]{revtex4-1}
\usepackage[latin9]{inputenc}
\usepackage[a4paper]{geometry}
\usepackage{xcolor}
\usepackage{pdfcolmk}
\usepackage{graphicx}
\usepackage{babel}
\usepackage{mathrsfs}
\usepackage{amsmath}
\usepackage{amssymb}
\usepackage{esint}
\PassOptionsToPackage{normalem}{ulem}
\usepackage{ulem}
\usepackage{dsfont} % for \mathds{1}
\usepackage[unicode=true,pdfusetitle,
 bookmarks=true,bookmarksnumbered=false,bookmarksopen=false,
 breaklinks=false,pdfborder={0 0 1},backref=false,colorlinks=false]
 {hyperref}

\makeatletter
\newcommand{\Rmnum}[1]{\expandafter\@slowromancap\romannumeral #1@}

\usepackage{bbold}

\let\textquotedbl="

\makeatother

\begin{document}

\title{Noisy independent component analysis of auto-correlated components }

\author{Jakob Knollm\"uller, Torsten A. En\ss lin}

\affiliation{{\small{}Max-Planck-Institut f\"ur Astrophysik, Karl-Schwarzschildstr.~1,
85748 Garching, Germany}\\
Ludwig-Maximilians-Universit\"at M\"unchen, Geschwister-Scholl-Platz{\small{}~}1,
80539 Munich, Germany}
\begin{abstract}

We present a new method for the separation of  superimposed, independent, auto-correlated components from noisy multi-channel measurement. The presented method simultaneously reconstructs and separates the components, taking all channels into account and thereby increases the effective signal-to-noise ratio considerably, allowing separations even in the high noise regime. Characteristics of the measurement instruments can be included, allowing for application in complex measurement situations. Independent posterior samples can be provided, permitting error estimates on all desired quantities. Using the concept of information field theory, the algorithm is not restricted to any dimensionality of the underlying space  or discretization scheme thereof.
\end{abstract}

\maketitle
 
\section{Introduction}
The separation of independent sources in multi-channel measurements is a fundamental challenge in a large variety of different contexts in the fields of science and technology. Large interest in such methods comes from bio-medicine, namely neural-science to investigate brain activities \citep{EEG}, but also in the analysis of financial time series \citep{finance} or for the separation of astrophysical components in our universe \citep{CMB}, to name a few.

Mainly two distinct approaches to component separation exist, namely Principle Component Analysis (PCA) and Independent Component Analysis (ICA).

PCA performs a linear transformation of the data to obtain mutually uncorrelated, orthogonal directions, which one calls the principle components. For  different principle  components $s_1$ and $s_2$ their covariance vanishes if averaged over the data set:
\begin{align}
\label{eq:uncorrelated}
\langle s_1 s_2 \rangle - \langle s_1\rangle \langle s_2 \rangle = 0
\end{align}
PCA is very useful in situations the data can be described by orthogonal processes. However, this does not imply independence, therefore higher order correlations may not vanish \citep{ICAAA}.
The number of principle components one obtains depends on the dimension of the involved data spaces. Some of these components are due to processes generating the data, others might just be due to noise. 
Drawing a line between those classes of components requires careful consideration of the context the data was obtained in.

ICA speaks of independent components $s_1$ and $s_2$ if and only if their probability distributions factorize
\begin{align}
\label{eq:independence}
\mathcal{P}(s_1,s_2) = \mathcal{P}(s_1)\mathcal{P}(s_2)
\end{align}
ICA Algorithms try to estimate independent components by maximizing some measure of independence. Several such measures are used, such as kurtosis, negentropy, or mutual information, to name a view. These all rely on the non-Gaussian statistics of the components. A mixture of Gaussian components is still Gaussian and does not have the non-orthogonal, relevant directions used in traditional ICA.  Therefore, it us often assumed that non-Gaussianity is a prerequisite of ICA.
However, the exploitation of  auto-correlations in a temporal or spatial domain breaks this Gaussian symmetry and allows identification of the components \citep{AMUSE}. 

An example for a PCA method which can be used in a rather similar setting to the one discussed in this work is the multivariate singular spectrum analysis (MSSA)\citep{MSSA}. 
It can also be used in noisy multi-channel measurement situations, taking auto-correlations into account. 
This is done by extending the original channel measurements with a number of time-delayed versions of the vectors. Then one calculates the correlation matrix of all possible channels and time delays. 
Diagonalizing this matrix leads to the orthogonal principle components, incorporating temporal correlations via the time delays. The most relevant principle components can then be used to describe main features of the data, allowing to analyze dynamical properties of the underlying system. 

We, however, want to identify the truly independent components in the data, as characterized by Eq. \ref{eq:independence}. For this goal a PCA method based on the weaker criteria of Eq. \ref{eq:uncorrelated} is a suboptimal approach.

On the side of Independent Component Analysis we have a large variety of widely used algorithms, the more popular ones include FastICA \citep{FastICA} and JADE \citep{JADE}, which rely on the above mentioned independence measures in noise free environments. The often inherent temporal or spatial correlation of the individual components is also not used. An algorithm which uses them is AMUSE \citep{AMUSE} which exploits time structure in a noise-free scenario.

A problem for ICA methods is often the presence of measurement noise. The noise prohibits a unique recovery of the individual components and demands for a probabilistic describtion of the problem. Several approaches have been made to solve this problem by using maximum likelihood methods \citep{MLICA} or Gaussian mixtures \citep{MOG}. In essence this method will follow a similar path.

The general advice in the literature so far, however, is to first de-noise the measurement and then to treat the results as noiseless, processing then with suitbale ICA methods \cite{ICAbook}. This approach severely suffers in the high noise regime as it is limited to the signal-to-noise ratio of the individual measurements.

The method we want to present combines the concept of auto-correlation with noisy measurements and thereby overcomes this restriction by reconstructing and separating the components simultaneously, combining the information across all measured channels and thereby vastly increases the effective signal-to-noise ratio while taking spatial or temporal correlations of the individual components into account. Using this method we can improve the result by adding additional channels and satisfying results are obtained even in high noise environments.

We achieve this by following the Bayesian framework to consistently include auto-correlations to a posterior estimate on the components. The posterior, however, is not accessible analytically and the maximum posterior estimate is insufficient for this problem. We will therefore present an approximation to the true posterior which is capable of capturing its essential features. We will use the Kullback-Leibler divergence to optimally estimate the model parameters in an information theoretical context. 

Furthermore we will formulate the components as physical fields without the requirement of specifying any discretization. This allows us to use the language of information field theory (IFT) \citep{IFT} to develop an abstract algorithm free of any limitations to be used in a specific grid or on a specific number of dimension.

 IFT is information theory for fields, generalizing the concept of probability distributions of functions over continuous spaces. In this framework we can formulate a prior distribution encoding the auto-correlation of the components. 

Fist we describe the generic problem of noisy independent component analysis. In the next section we formulate auto-correlations in continuous spaces and how to include them in our model. Ways how to approximate the model in a feasible fashion are discussed in Sec.\,\ref{sec:approximating the posterior}. In order to infer all parameters we have to draw samples from the approximate posterior. We describe a procedure how to obtain such samples. After briefly stating the full algorithm, we discuss its convergence and demonstrate its performance on two numerical examples showcasing different measurement situations.

\section{Noisy ICA}
The noisy ICA \citep{noisyICA} describes the situation of multiple measurements of the same components in different mixtures in the presence of noise.
 Each individual measurement $i$ at some position or time $x$ results in data $d_{i,x}$ which has a noise contribution $n_{i,x}$, as well as a linear combination $M_{ij}$ of all components $s_{j,x}$ which also have some spatial or temporal structure. The data equation for this is process is given by
 \begin{align}
d_{i,x} = M_{ij} s_{j,x} +n_{i,x} \text{ .}
\end{align}
Here we use the summation convention over multiple indices. The mixture $M_{ij}$ acts on all positions equally and therefore does not depend on a position index. We can simplify the notation of the equation above by simply dropping the position index, interpreting those quantities as vectors. What remains are the measurement and component indices.
\begin{align}
d_i = M_{ij} s_j +n_i
\end{align}
We can go even further by introducing the multi-measurement vector $d$ as a vector of vectors, containing the individual measurements, and noise $n$, as well as the multi-component vector $s$, consisting of all components. Then one can use the usual matrix multiplication with the mixture $M$ to end up with the index-free formulation of this equation as 
\begin{align}
d = M s + n \: \text{.}
\end{align}

We want to modify this expression in two ways. The first one is to describe the components not as vectors, but as fields. On the one hand  true components usually should resemble some physical reality, which is not limited to any discretization and therefore best described by a continuous field, therefore
\begin{align}
s_{j,x} \rightarrow s_j(x) \text{\: .}
\end{align}
On the other hand our data $d$ can never be a continuous field with infinite resolution, as this would correspond to an infinite amount of information. It is therefore necessary to introduce a description of the measurement process, where some kind of instrument probes the physical reality in form of the mixed continuous components. In general this instrument is a linear operator with a continuous physical domain and discrete target, the data space.  Including this response operator $R$, the data equation becomes
\begin{align}
d_{i,X} = \int dx \: R_i(X,x) M_{ij} s_j(x) + n_{i,X} \text{ .}
\end{align} 
The capital letter $X$ represents the discrete positions of the data, whereas $x$ is the continuous position. We can again drop all indices and state the equation above in operator notation
\begin{align}
d =R M s + n \text{\: .}
\end{align}
We have now decoupled the domains of the data from the components. We can also not represent components with an infinite resolution, once we want to do numerical calculations we have to somehow specify a discretization, but introducing the response operator allows us to choose representations completely independent from the data and the measurement process. The response operator also allows us to consider any linear measurements, using different instruments for the individual channels. One can easily include masking operations, convolutions, transformations or any other linear instrument specific characteristics in a consistent way.

We will now derive the likelihood of this data model.
The noise $n$ will be assumed to be Gaussian with known covariance $N$ and vanishing mean in the data domain. We describe it as 
\begin{align}
\label{eq:noiseprior}
\mathcal{P}(n) = \mathcal{G}(n,N) = \frac{1}{\vert 2 \pi N \vert^\frac{1}{2}} e^{-\frac{1}{2} n^\dagger N^{-1} n} \text{ .}
\end{align}
The expression $n^\dagger$ is the complex conjugated, transposed noise vector. This  leads to a scalar in the exponent via matrix multiplication.
Using this data equation, we can derive the likelihood of the data $d$, given components $s$, mixture $M$ and noise realization $n$. This is a delta distribution as the data is fully determined by the given quantities. 
\begin{align}
\mathcal{P}(d\vert s,M,n) =&\: \delta(d-(RMs+n)) 
\end{align}
However, the realization of the noise is not of interest and we will marginalize it out using the Gaussian noise model given in Eq. \ref{eq:noiseprior}.
\begin{align}
\mathcal{P}(d\vert s,M) =& \int \mathcal{D}n \:\delta(d-(RMs+n)) \mathcal{G}(n,N)\\
=& \: \mathcal{G}(d-RMs,N)
\end{align}
Taking the negative logarithm provides us with the information Hamiltonian\footnote{
The information Hamiltonian emerges from the analogy (or equivalence) of information theory to statistical physics:
\begin{align}
\mathcal{P}(s\vert d) =& \frac{\mathcal{P}(s,d)}{\mathcal{P}(d)} \equiv \: \frac{e^{-\mathcal{H}(s,d)}}{\mathcal{Z}(d)} \text{, with information Hamiltonian} \\
\mathcal{H}(s,d) =& - \mathrm{ln} \mathcal{P}(s, d) \text{and partition function} \\ 
\mathcal{Z}(d) =& \int Ds \: e^{- \mathcal{H}(s,d)} 
\end{align}
 $\mathcal{H}(s,d)$ therefore contains all available information on the signal $s$ and is often a more practical object perform do calculations with than the equivalent probability distributions, as information Hamiltonians are additive:
 \begin{align}
 \mathcal{H}(s,d) = \mathcal{H}(d\vert s) + \mathcal{H}(s)
 \end{align}
}
of the likelihood, also called negative log-likelihood.
\begin{align}
\mathcal{H}(d\vert s,M) \equiv& \: - \mathrm{ln}\left[\mathcal{P}(d\vert s,M)\right]\nonumber \\
 = & \:\frac{1}{2} (d-RMs)^\dagger N^{-1} (d-RMs) \nonumber \\
 &+\frac{1}{2} \mathrm{ln}\vert 2 \pi N \vert
\end{align}

\section{auto-correlation}
The components we want to separate exhibit auto-correlation and we want to exploit this essential property. A component $s_i(x)$ has some value at each location in its continuous domain. We define the scalar product for two fields as $ j^\dagger s \equiv \: \int dx \: j^*(x) s(x)$, where $j^*(x)$ expresses the complex conjugate of the field $j$ at position $x$. We can express the two-point auto-correlation as
\begin{align}
S_i(x,x') \equiv \: \langle s_i(x) s_i^*(x')\rangle_{\mathcal{P}(s_i)} \text{ ,}
\end{align}
which is a linear operator encoding the internal correlation of the component $s_i$.
Assuming statistical homogeneity, the correlation between two locations $S_i(x,x')$ only depends on their position relative to each other.
\begin{align}
S_i(x,x')=S_i(x-x')
\end{align}
Furthermore, we can now apply the Wiener-Khinchin theorem \citep{Khintchin} and identify the eigenbasis of the correlation with the associated harmonic domain, which for flat spaces corresponds to the Fourier basis. This is convenient for the implementation of the algorithm because it allows us to apply the correlation operator in Fourier space, where it is just a diagonal operation and efficient implementations for the Fourier transformation of the components are available as well. This approach stays feasible even for high resolutions of the components, as the representation of the covariance scales roughly linearly in Fourier space, but quadratically in position space. 

For components with correlations in more than one dimension, it might also be advantageous to assume statistical isotropy. With this, the correlation only depends on the absolute value of the distance between two points. We can then express the correlation structure by a one-dimensional power spectrum.

In this paper we assume the correlation structure of a component to be known.
 In principle it could also be inferred from the data with critical filtering \citep{smoothpower}. 
 The idea of critical filtering is to parametrize the power spectrum and additionally infer its parameters. 
 This allows us to separate auto-correlated components without knowing the correlation structure beforehand. Critical filtering has been successfully applied in multiple applications \citep{CFGEO, RESOLVE, D3PO} and can be included straightforwardly in this model. 
 In order to keep the model simple we choose not to discuss this case in detail here.

We use the known correlation structure $S_i$ to construct a prior distribution of the components $s_i$, informing the algorithm about the auto-correlation. The least informative prior with this property will be a Gaussian prior with vanishing mean and covariance $S_i$
\begin{align}
\mathcal{P}(s_i) = \mathcal{G}(s_i,S_i) \:\text{.}
\end{align}
Conceptually this is a Gaussian distribution over the continuous field $s_i$. In any numerical application we have to represent the field in a discretized way and this distribution becomes a regular multivariate Gaussian distribution again. 
Assuming independence of the individual components, the prior distributions factorize and we write 
\begin{align}
\mathcal{P}(s) = &\prod_i \mathcal{G}(s_i,S_i) \\
\equiv & \: \mathcal{G}(s,S) \text{ .}
\end{align}
The product of Gaussian distributions can be written in the compact form of a combined Gaussian over the multi-component vector $s$ with block-diagonal correlation structure expressing the independence of the different components of each other, i.e. $\langle s_i(x)s_j(x') \rangle_{\mathcal{P}(s)} = 0$ for $i \neq j$. The prior independence actually implements the underlying assumption of any ICA method as stated in Eq. \ref{eq:independence}.

It is worth emphasizing that this way of formulating the correlation structure allows us to apply the resulting algorithm regardless of the dimension. At the end we will demonstrate one dimensional cases for illustration purposes, but without any changes the algorithm generalizes to two, three and n-dimensional situations. Even correlations on curved spaces such as on a sphere can be considered by replacing the Fourier basis with the corresponding harmonic basis.

The information Hamiltonian of this prior distribution is given by 
\begin{align}
\mathcal{H}(s) = \frac{1}{2} s^\dagger S^{-1} s + \frac{1}{2} \mathrm{ln}\vert 2\pi S \vert \: \text{.}
\end{align}
We now constructed a likelihood from our data model and a prior distribution over the components, encoding their auto-correlation. Using Bayes theorem we can derive the posterior distribution over the components $s$ and their mixture $M$ via
\begin{align}
\mathcal{P}(s,M\vert d) = \frac{\mathcal{P}(d\vert s, M) \mathcal{P}(s, M)}{\mathcal{P}(d)} \: \text{.}
\end{align}
We did not discuss any prior distribution over the mixture $M$ as we do not want to restrict it in any way. Any problem-specific insights about the mixture should be expressed right here. The prior distributions can in our case be written as 
\begin{align}
\mathcal{P}(s, M) \propto \mathcal{P}(s) = \mathcal{G}(s,S)
\end{align}
and thereby implicitly assuming a flat and independent priors on the entries of $M$.
The evaluation of $\mathcal{P}(d)$ is not feasible as it involves the integration over both, the mixture and signal  of the joint probability distribution. Therefore we have to think of approximative approaches.
First we state the posterior information Hamiltonian $\mathcal{H}(s,M\vert d) = -\mathrm{ln}(\mathcal{P}(s,M\vert d))$ without any component or mixture independent terms.
\begin{align}
\label{eq:hamiltonian}
\mathcal{H}(s,M\vert d) = &\frac{1}{2} s^\dagger M^\dagger R^\dagger N^{-1} RMs -s^\dagger M^\dagger R^\dagger N^{-1} d \nonumber \\&+ \frac{1}{2} s^\dagger S^{-1} s + \mathrm{const}(d)\mathrm{.}
\end{align}

\section{approximating the posterior}
\label{sec:approximating the posterior}
\begin{figure}

\includegraphics[scale=0.56]{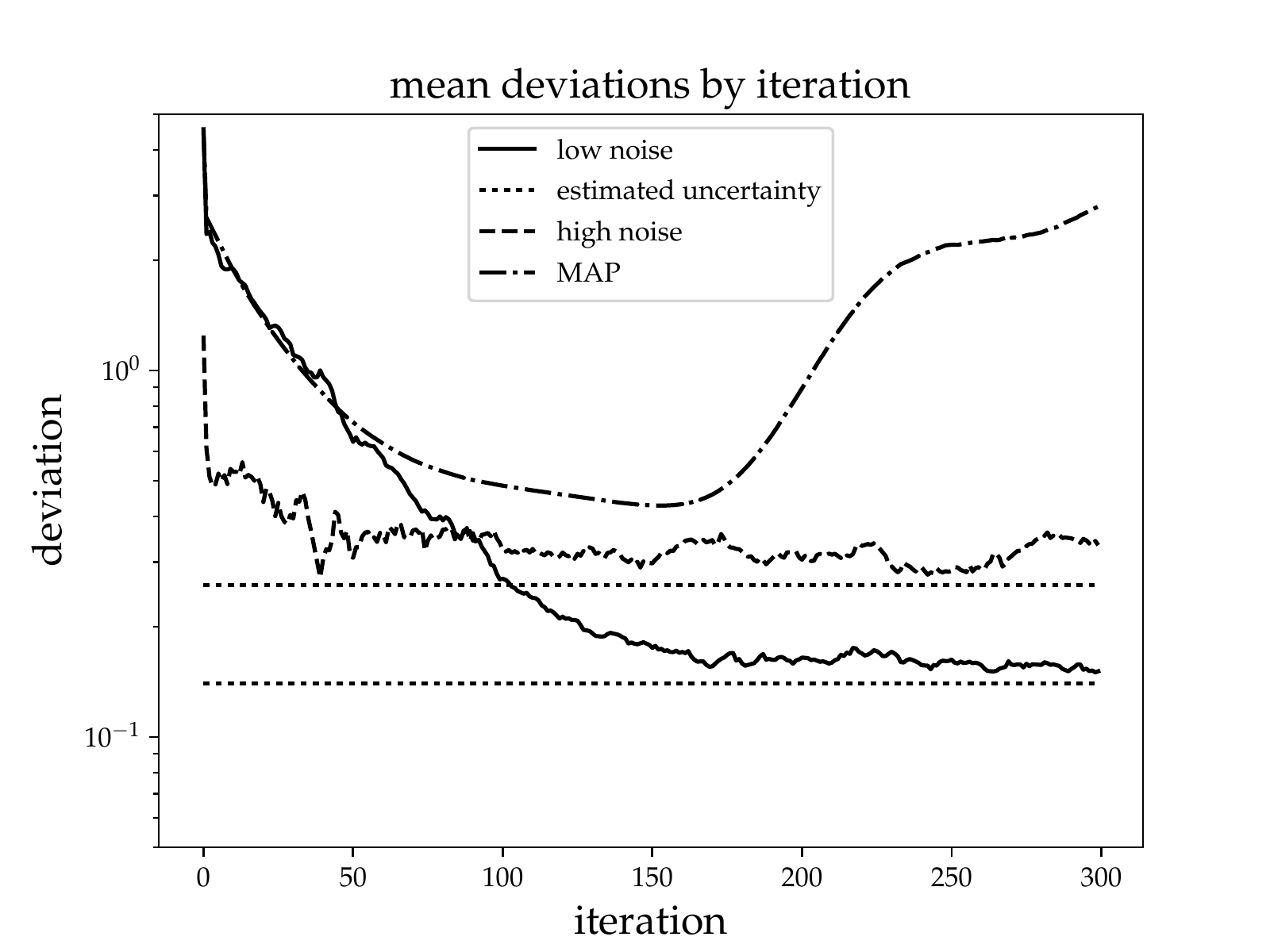}
\caption{The mean deviation of the current estimates from the true components during the minimization for all three example scenarios compared to the estimated uncertainty of the final result.}
\label{fig:convergence}
\end{figure}

A typical approach to a problem like this is to take the most likely posterior value as an estimate of the parameters. This is achieved by minimizing the information Hamiltonian above. It can be interpreted as an approximation of the posterior distribution with delta distributions peaked at the most informative position in the sense of minimal Kullback-Leibler (KL) divergence \citep{KLdivergence} between true and approximated posterior. For this the latter can be written as
\begin{align}
\widetilde{\mathcal{P}}_{\mathrm{MAP}}(s,M\vert d) = \delta(s-s_{\mathrm{MAP}})\delta(M-M_{\mathrm{MAP}}) \:\text{.}
\end{align}
This approximation turned out to be insufficient for a meaningful separation of the components as we will illustrate in Sect. \ref{sec:examples}.
 Iterating the minimization with respect to the components and the mixture we do not obtain satisfying results. The maximum posterior estimate is known to over-fit noise features. This has severe consequences in this component separation as it relies on iterative minimization of the Hamiltonian with respect to one of the parameters. In each step we over-fit which affects the consecutive minimization. In this way we accumulate errors in the parameters, leading to unrecognizable, strongly correlated components. During the minimization the MAP algorithm approaches reasonable component separations but it does not converge to those and continues to accumulate errors, converging somewhere else. This behavior can be seen in Figure \ref{fig:convergence}, showing the deviation of current estimates to the true components for the MAP case, as well as the algorithm discussed in the following.

Our strategy to solve this problem is to choose a richer model to approximate the posterior distribution which is capable to capture uncertainty features and reduce over-fitting. Instead of using a delta-distribution to describe the posterior components, we choose a variational approach using  a Gaussian distribution whose parameters have to be estimated. For the posterior mixture we stay with the initial description of the point estimate as this turns out to be sufficient for many applications.
We therefore approximate the true posterior with a distribution of the form
\begin{align}
\label{eq:approximation}
\widetilde{\mathcal{P}}(s,M\vert d) = \mathcal{G}(s-m,D)\delta(M-M_*) \: \text{.}
\end{align}
In this approximation we describe our posterior knowledge about the mixture $M$ with the point-estimate $M_*$ and about the  components $s$ by a Gaussian distribution with mean $m$ and covariance $D$ . We will use $m$ as the estimate of our posterior components and the covariance $D$ describes the uncertainty structure of this estimate. Compared to the prior covariance $S$, the posterior covariance does not have to be diagonal in the harmonic domain, as the likelihood typically breaks the homogeneity.

The main problem of this approximation is the point-estimate of the posterior mixture $M$. With this we assume perfect knowledge about the mixture with absolute certainty. This is certainly not justified due to the probabilistic nature of the problem, but this is true for every point-estimate in any context. 
This approximation also affects the posterior covariance of the components $D$ which will contain the mixture. As we assume no uncertainty in it, we will not consider any errors in the mixture and therefore underestimate the true uncertainty of the components. In the low noise regime this effect is negligible, it will become larger for low signal-to-noise ratio, as we will see in the numerical examples.
This model, however, seems to perform reasonably well in relatively high noise regimes, but one has to take the error estimates with caution and keep in mind that those will be underestimated. 
One could easily think of a more complex model performing even better and more accurate in the high noise case.
For example also approximating the mixture with a Gaussian distribution or using one large Gaussian distribution, also accounting for cross-correlations between components and the mixture.
Those models come with the cost of dramatically increased analytical and numerical complexity. As no analytic form of the posterior is available, the best solution possible can be obtained from sampling the posterior, which can can become computationally extremely expensive, as the  dimensionality of the problem scales with the resolution of the components.
We choose the approximation given in Eq. \ref{eq:approximation} as it should capture the relevant quantities while being as simple as possible.

In order to estimate the parameters of the distribution in Eq. \ref{eq:approximation}, we have to minimize its  KL divergence to the initial posterior. The divergence is defined as
\begin{align}
\mathrm{KL}&\left[\widetilde{\mathcal{P}}(s,M\vert d)\vert\vert\mathcal{P}(s,M\vert d)\right] \equiv\\
\equiv &\int \mathcal{D}s \:\mathcal{D}M\: \widetilde{\mathcal{P}}(s,M\vert d) \:\mathrm{ln}\left[\frac{\mathcal{P}(s,M\vert d)}{\widetilde{\mathcal{P}}(s,M\vert d)}\right]\\
= &  \:\left\langle \mathcal{H}(s,M_{*}\vert d)\right\rangle_{\mathcal{G}(s-m,D)} \nonumber  \\
&- \left\langle \mathrm{ln}\left[\mathcal{G}(s-m,D)\right]\right\rangle_{\mathcal{G}(s-m,D)} \text{ .}
\end{align}
The integration over the mixture just replaces every $M$ by $M_*$. In order to keep the expressions shorter we will drop from now on the star and will use in all further calculations just the symbol $M$.  We now have to calculate Gaussian expectation values of the total information Hamiltonian. We can perform this calculation with the cyclical property of the trace operation and the identity 
\begin{align}
\langle s s^\dagger\rangle_{\mathcal{G}(s-m,D)} = mm^\dagger + D \:\text{.}
\end{align} 
The second expectation value in the KL-divergence corresponds to the entropy of the Gaussian distribution. The analytic expression then reads
\begin{align}
\label{eq:KL}
\mathrm{KL} = &\: \frac{1}{2} m^\dagger M^\dagger R^\dagger N^{-1} R M m + \frac{1}{2} \mathrm{Tr}\left[M^\dagger R^\dagger N^{-1}RMD\right]\nonumber \\
&\: - m^\dagger M^\dagger R^\dagger N^{-1} d \nonumber \\
& \: +\frac{1}{2} m^\dagger S^{-1} m +\frac{1}{2}\mathrm{Tr}\left[S^{-1}D\right] \nonumber\\
& \: +\mathrm{Tr}\left[ 1 +\mathrm{ln}(2\pi D)\right] \:\text{.}
\end{align}
We have to minimize this expression with respect to all parameters of our approximate distribution, namely $m$, $D$ and $M$. 

We will start with the posterior component mean $m$. Comparing the terms of the Hamiltonian in Eq. \ref{eq:hamiltonian} containing $s$ with the ones in the KL containing $m$ we find their analogous structure. Given some mixture $M$ the minimum will be identical. We can solve for the posterior mean by setting the derivative of the KL-divergence with respect to it to zero:
\begin{align}
\frac{\delta \mathrm{KL}}{\delta m^\dagger} \overset{!}{=}& \:0\\ = & \:-M^\dagger R^\dagger N^{-1}d\nonumber\\& +  M^\dagger R^\dagger N^{-1} R M m + S^{-1} m\\
\Rightarrow m = & \: \left(M^\dagger R^\dagger N^{-1}RM + S^{-1}\right)^{-1}M^\dagger R^\dagger N^{-1}d
\end{align}
The structure of the solution looks familiar. In fact it is the Wiener filter solution \citep{Wiener} for a known mixture. We can also solve for the posterior covariance:
\begin{align}
\frac{\delta \mathrm{KL}}{\delta D} \overset{!}{=}& \: 0 \\ 
\Rightarrow D =& \: \left(M^\dagger R^\dagger N^{-1} R M + S^{-1} \right)^{-1}
\end{align}
This also turns out to be the Wiener covariance for known mixture. We can then define the information source $j$ and write the approximate posterior mean in terms of the Wiener filter formula:
\begin{align}
j \equiv & \:M^\dagger R^\dagger N^{-1} d \\
m =&\: Dj
\end{align}
If we know the mixture $M$ a Gaussian with mean $m$ and covariance $D$ would be the exact posterior of the components given the data. 
\\

Now we also have to calculate the derivative of the KL-divergence with respect to the entries of the mixture matrix $M_{ij}$ while keeping $m$ and $D$ fixed.
In this calculation the trace term 

\begin{align}
\label{eq:corr}
\frac{1}{2} \mathrm{Tr}\left[M^\dagger R^\dagger N^{-1}RM D\right]
\end{align}
in the divergence does not vanish as it contains the mixture $M$ and will gives rise to the required uncertainty corrections, regularizing the mixture and therefore making the algorithm converge.
Unfortunately this term is numerically challenging. The trace of an operator can be extracted via operator probing  \citep{traces, diagonals}.
This involves multiple numerical operator inversions using  conjugate gradient method which is computationally expensive.

We will choose another approach which avoids the trace expressions by taking them implicitly into account. For this purpose we still have to solve multiple linear systems, but we found the new approach to be numerically more stable and more general as it can also be applied in cases we do not have explicit expressions.

In order to obtain the analytic expression of the KL-divergence we calculated the expectation value of the information Hamiltonian with respect to the approximate posterior Gaussian distribution which gave rise to the trace terms in the first place. 
To avoid them we will consider the KL-divergence before performing the averaging over the approximating Gaussian and keep it that way during the derivation of the gradient. 
To estimate the resulting expressions we approximate the averaging by replacing it with an average over samples drawn from the distribution $\mathcal{G}(s-m,D)$.
All relevant terms in the KL-divergence concerning $M$ read:
\begin{align}
\label{eq:avrg}
\mathrm{KL}_{M} = \frac{1}{2}\langle s^\dagger M^\dagger R^\dagger N^{-1} RMs\rangle_{\mathcal{G}(s-m,D)} \nonumber\\
 - \langle d^\dagger N^{-1} RMs \rangle_{\mathcal{G}(s-m,D)}
\end{align}
For the minimization with respect to the mixture $M$ we assume the posterior mean $m$ and covariance $D$ to be fixed, so we can calculate the derivative of the expression above with respect to the mixture ignoring the expectation value:
\begin{align}
\frac{\delta \mathrm{KL_M}(s,M\vert d)}{\delta M_{ij}} = \langle s^\dagger M^\dagger R^\dagger N^{-1} R\mathbb{1}_{ij}s\rangle_{\mathcal{G}(s-m,D)} \nonumber\\
 - \langle d^\dagger N^{-1} R\mathbb{1}_{ij}s \rangle_{\mathcal{G}(s-m,D)}
\end{align}
The operator $\mathbb{1}_{ij}$ with $\left(\mathbb{1}_{ij}\right)_{i'j'} = \delta_{ii'}\delta_{jj'}$ singles out the position of the entry $ij$. It is of the same shape as the mixture matrix with all entries zero, except for the one at the position $ij$.
Comparing this term to the derivative of the information Hamiltonian 
\begin{align}
\frac{\delta \mathrm{H}(s,M\vert d)}{\delta M_{ij}} =&\: s^\dagger M^\dagger R^\dagger N^{-1}R\mathbb{1}_{ij}s \nonumber \\
& - d^\dagger N^{-1} R \mathbb{1}_{ij} s
\end{align}
which is used in the maximum posterior approximation, the main difference to our method becomes apparent. In the maximum posterior approach only a point estimate for the components $s$ is used. Our approach replaces the minimization of the Hamiltonian with the minimization of the mean Hamiltonian under the approximated Gaussian, taking the uncertainty structure of the components into account.

Setting the our mixture gradient to zero allows us to solve for the mixture in a Wiener-filter-like fashion

\begin{align}
M = \left\langle s^\dagger \mathbb{1}^\dagger R^\dagger N^{-1}R\mathbb{1}s\right\rangle^{-1} \left\langle d^\dagger N^{-1}R\mathbb{1}s\right\rangle
\end{align}
The first part serves as a Wiener covariance and the second term corresponds to an information source for $M$. 

At some point we have to evaluate all those expectation values numerically to minimize the divergence with respect to the mixture $M$. 

The terms we want to calculate are expectation values of the Gaussian distribution $\mathcal{G}(s-m,D)$, but they will introduce impractical trace terms. Instead we want to approximate it with a set of $L$ samples $\{s^*\}$  distributed according to $\mathcal{G}(s-m,D)$ using the sampling distribution.
\begin{align}
\mathcal{G}(s-m,D) \approx \frac{1}{L} \sum_{l=0}^L \delta( s - s^*_l)
\end{align}
 Using this distribution, the expectation values are replaced with the average over the set of samples. 
 In the next section we will discuss how to obtain those samples from the distribution.
 
\section{Approximate posterior sampling}
Drawing samples from the approximate posterior distribution for the components is challenging, as we do not have direct access to its eigenbasis in which the correlation structure is diagonal.
If we had, we could draw independent Gaussian samples with mean zero and variance one in each dimension, weight them with the  square root of the eigenvalue to adjust to the correct variance and apply the transformation to position space given by the eigenvectors. 
At this point the sample has the correct correlation structure and has only be adjusted to the correct mean by adding it.
 
The main task is therefore to get samples with the correct correlation structure. In the case of our approximate posterior
\begin{align}
\widetilde{\mathcal{P}}(s\vert d) = \mathcal{G}(s-m,D) \text{,}
\end{align} 
we have to find residuals $(s-m)$ which satisfy
\begin{align}
\langle (s-m) (s-m)^\dagger \rangle_{\mathcal{G}(s-m,D)}= D \text{ .}
\end{align}
Obviously we do not have access to the true components $s$. What we do have is a prior belief about them. Its correlation is diagonal in the Fourier domain for each component and we can easily generate a samples from it using the description above.
\begin{align}
s' \curvearrowleft \mathcal{G}(s,S)
\end{align}
Those components $s'$ have nothing to do with the true components, except their correlation structure. We now want to find an $m'$ which satisfies
\begin{align}
\langle (s'-m') (s'-m')^\dagger \rangle_{\mathcal{G}(s'-m',D)}= D \text{ .}
\end{align}
The posterior covariance is a Wiener Filter covariance described by 
\begin{align}
D^{-1} = M^\dagger R^\dagger N^{-1} R M + S^{-1}
\end{align}
with given mixture $M$, instrument response $R$, noise covariance $N$ and prior signal covariance $S$.
We can reconstruct the quantity $m'$ from the data we would have obtained if $s'$ were the real components. 
We therefore have to simulate the measurement process of the arbitrary sample $s'$ using our linear data equation
\begin{align}
d' = RMs' + n' \text{ .}
\end{align}
We can draw an noise realization $n'$ from the prior noise distribution $\mathcal{G}(n,N)$ which is diagonal in data space. On this mock data we simply perform a Wiener Filter reconstruction
\begin{align}
m' = D j' \: \text{, with}\\
j' \equiv \: M^\dagger R^\dagger N^{-1} d' \text{ .}
\end{align}
This is the numerical costly part of the sampling procedure as it involves a conjugate gradient to solve the system.

However, once we obtained the reconstruction $m'$ of the mock signal $s'$ we can calculate the residual, which follows exactly the correlation structure encoded in $D$. The components $s'$ are now a sample drawn from the distribution $\mathcal{G}(s' - m', D)$. What we actually want is a sample $s^*$ from $\mathcal{G}(s^* -m, D)$, originating from our true data. The residuals of both distributions have the same statistical properties, thus we can therefore set them equal and solve for $s^*$.

\begin{align}
s' - m' \overset{!}{=} s^* -m \\
s^* = s' -m' +m
\end{align}
Those components $s^*$ now exactly behave according to $\mathcal{G}(s^* -m, D)$ with mean $m$ and covariance $D$. We can use samples drawn by this procedure to calculate the expectation values we need in the minimization process for the mixture.

Furthermore,  we can use the samples to easily estimate arbitrary posterior properties, such as the uncertainty of our component estimate.

Let us briefly summarize this approach for approximate posterior sampling. We start with a sample $s'$  drawn independently from the component prior, use those to set up a mock observation, which  provides us with mock data $d'$. We Wiener filter this data to get the posterior mean $m'$. The only thing we are interested in from this calculations is the residual $s'-m'$, as it allows us to construct a sample $s^*$ from the mean $m$ of the distribution we are actually interested in.
The more samples we draw this way the  better the sampling distribution approximates the true distribution. However, we want to use as few samples as possible as their calculation is computationally expensive, not only during the sampling procedure, but also their usage in all further calculations, such as gradient estimations. During the alternating minimization with respect to the mean components $m$ and the mixture $M$, we have to permanently recalculate the samples as the mean and mixture is constantly changing. We found that it is practical to start with few samples and to increase their number during the inference. Note that the KL divergence is not fully calculated and also only estimated through the samples, therefore this estimate inherits stochastic variations.

\section{The Algorithm}
Now we have all the tools to set up an iterative scheme to minimize the KL divergence in order to infer the parameters of our approximation.

In order to use this algorithm we need knowledge on the characteristic noise behavior encoded in the correlation structure $N$, as well as on the statistical properties of the individual components described by the prior covariance $S$.
In addition we have to specify the number of components we want to infer. 

We will start with a random guess for the mixture $M$ and use it to estimate our initial  mean components $m$ and covariance $D$ under the assumption the initial guess of the mixture is correct using the Wiener Filter.
\begin{align}
D^{-1}& = M^\dagger R^\dagger N^{-1} R M + S^{-1} \\
j &=\: M^\dagger R^\dagger N^{-1} d \\
m &=\: Dj
\end{align}

We have now the first estimate of the approximate posterior distribution $\mathcal{G}(s-m, D)$. In order to estimate a new mixture we can draw a set of independent samples $\{s^*\}$ from this distribution using the procedure described in the previous section.
\begin{align}
 \{s^*\} \curvearrowleft \mathcal{G}(s-m,D) 
\end{align}
We use those to replace the Gaussian expectation values  with averages over the sampling distribution, which allows us to solve for a new estimate for the mixture, using the Wiener-Filter-like formula: 
\begin{align}
M = \left\langle s^\dagger \mathbb{1}^\dagger R^\dagger N^{-1}R\mathbb{1}s\right\rangle^{-1}_{\{s^*\}} \left\langle d^\dagger N^{-1}R\mathbb{1}s\right\rangle_{\{s^*\}}
\end{align}

Now we have to take care of the multiplicative degeneracy between between the components and their mixture vector. We therefore normalize each column of the mixture to an  $L_2$-norm of $\vert\vert M^\dagger_j\vert\vert=1$, multiplying each component mean accordingly by the normalization factor to keep the product $Mm$ unchanged.

If the power-spectrum of  the components are unknown, we perform here a critical filter step \citep{smoothpower}, which we choose not to discuss at this point.

This way we obtain a new estimate for the mixture, which allows us to estimate new component means and covariances, which allows us to draw new samples, which we can use for a new mixture, and so on, until the algorithm converges. We will discuss its converges in the next section. 

However, after the algorithm has converged we can use the samples to calculate any posterior quantity of interest involving the components and estimate its uncertainty. One example would be the spatial uncertainty of the component reconstruction  by evaluating 
\begin{align}
D_{xx} = \left\langle (s_x-m_x)^2  \right\rangle_{\{s^*\}} \: \text{.}
\end{align}

\section{on its convergence}
Each estimate of a new parameter on its own will reduce the remaining KL divergence between our approximated posterior and the true posterior, at least stochastically. The stochasticity is due to the noise introduced by the sampling and can be reduced by using more samples, for the price of high computational cost. Let us briefly discuss the symmetries, structure and minima of the Kullback-Leibler divergence as it is stated in Eq. \ref{eq:KL}. We start with two likelihood contributions
\begin{align}
\mathrm{KL} \: \widehat{=} \:\frac{1}{2} m^\dagger M^\dagger R^\dagger N^{-1} R M m  - m^\dagger M^\dagger R^\dagger N^{-1} d \text{.}
\end{align}
Here, we have a unique minimum for the mixed components $Mm$ due to the quadratic structure in the case $R^\dagger N^{-1} R$ is a full rank operator, otherwise its null space is unconstrained. 

In addition to this, individual mixtures $M$ and component means $m$, the terms above exhibit two symmetries, as we can multiply the mixtures for each components with arbitrary factors while dividing the corresponding components by according factors. This introduces a submanifold of minimal energy.

Finally, we can just interchange the components while also swapping the entries of the mixing matrix. Depending on the number of components, we get additional $c!$ times as many minima, with $c$ being the number of components. 

The only other terms concerning the mixture and components are their other likelihood contribution and the component prior term

\begin{align}
\mathrm{KL}  \: \widehat{=} \:\frac{1}{2} \mathrm{Tr}\left[M^\dagger R^\dagger N^{-1}RMD\right] +\frac{1}{2} m^\dagger S^{-1} m \text{\: ,}
\end{align}
which are both quadratic terms in $m$ and $M$, respectively, with positive sign and therefore do not introduce additional minima, but eliminate some degeneracy. First of all, these terms constrain the null space degeneracy to one single point. The multiplicative degeneracy between $M$ and $m$ is also broken as both quadratic terms regularize like $L_2$ norms. What remains  is the degeneracy of the multiplication of $M$ and $m$ with $-1$ for each component, allowing for $2^c$ possibilities. Thus, instead of entire submanifolds of optimal solutions we end up with a total of  $2^c c!$ minima of the KL divergence with respect to $M$ and $m$.

 In the case that the prior covariances for the individual components in $S$ are not identical, the interchange symmetry is broken and all those minima do not have the same divergence anymore. Therefore, using a gradient descent method we do not necessarily end up in a global minimum. This can be solved by discrete optimization steps, trying all possible permutations of the mixture and components and picking the one with smallest KL divergence.

This problem also vanishes if one also infers the prior correlation structure as the prior then adapts to the chosen permutation, leading to a global minimum for sure. 

We have seen that in the case of the same prior correlation structures for all components all minima of the divergence are global minima and we therefore will converge to an optimal solution irrespective of the starting position.

The speed of convergence, however, is hard to estimate as we rely on the iteration of consecutive minimizations of our parameters. Each individual minimization converges rather quickly, depending on the condition numbers of the matrices involved, as we invert them by the conjugate gradient method. The total convergence rate should depend on the correlation between the component means $m$ and mixtures $M$ in the KL divergence. The less they are correlated, the faster the individual parameters should reach a minimum. Strong correlations, however,  do not allow for large steps, therefore being slower.

Practically the computational effort highly depends on the choice of various quantities.
The algorithm is divided into two distinct minimizations for the mean components $m$ and mixture $M$ of different dimensionality, which is the main source of computational cost.
The dimensionality of the component part scales linearly with the number of components and their resolution, at least in the one-dimensional case. For higher dimensional components the resolution scales accordingly. The costly part in this minimization is the numerical inversion of an implicit operator in order to solve a Wiener Filter problem.
The minimization with respect to the mixture is rather cheap, having the dimension of number of components times number of data channels. Drawing one posterior samples, however requires a Wiener Filter of the complexity of the first part. We therefore want to keep the number of samples as low as possible, at least at the beginning of the inference. We can increase the number of samples towards the end, reducing the statistical sampling noise.

The entire algorithm consists of a large number of consecutive minimizations. The accuracy to which each of them is performed greatly effects the overall performance. We want to avoid unnecessary accuracy wherever possible, as all parameters are changing constantly and for the mixture the KL divergence is only a statistical estimate with uncertainties itself. Therefore we would waste computation if we aim for high accuracy initially. Towards the end, as the number of samples increases, one might also increase the accuracy. How to optimally steer this is rather difficult and currently requires case by case optimization.

\section{Numerical Examples}
\label{sec:examples}
\begin{figure}
\begin{center}

\includegraphics[scale = 0.56]{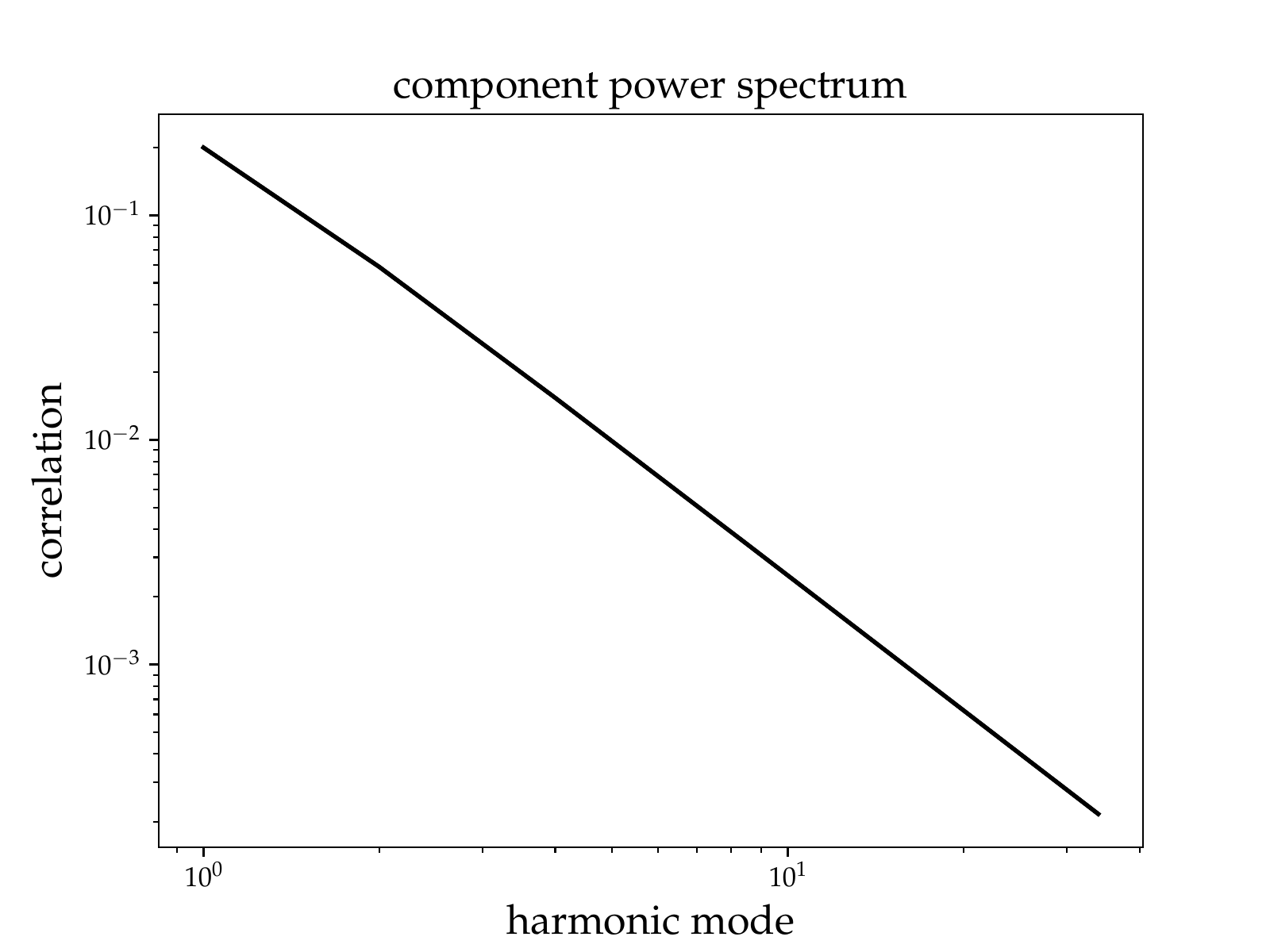}
\caption{The correlation structure of both components in Fourier space in double logarithmic representation. }
\label{fig:power}
\end{center}
\end{figure}

\begin{figure}
\begin{center}
\includegraphics[scale = 0.56]{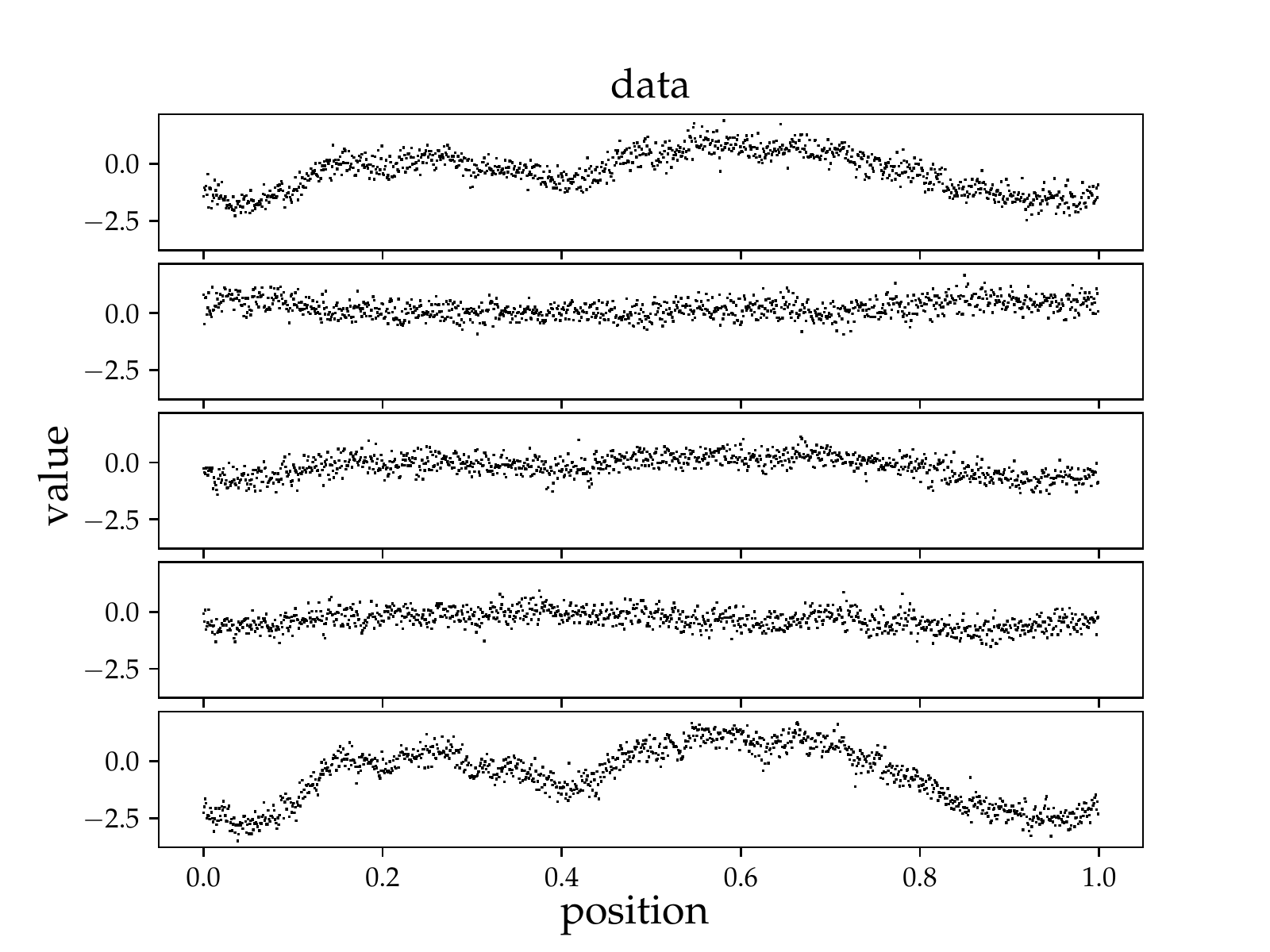}
\caption{Data of the first scenario in five channels from noisy measurements of two linearly mixed components.}
\label{fig:lowdata}
\includegraphics[scale = 0.56]{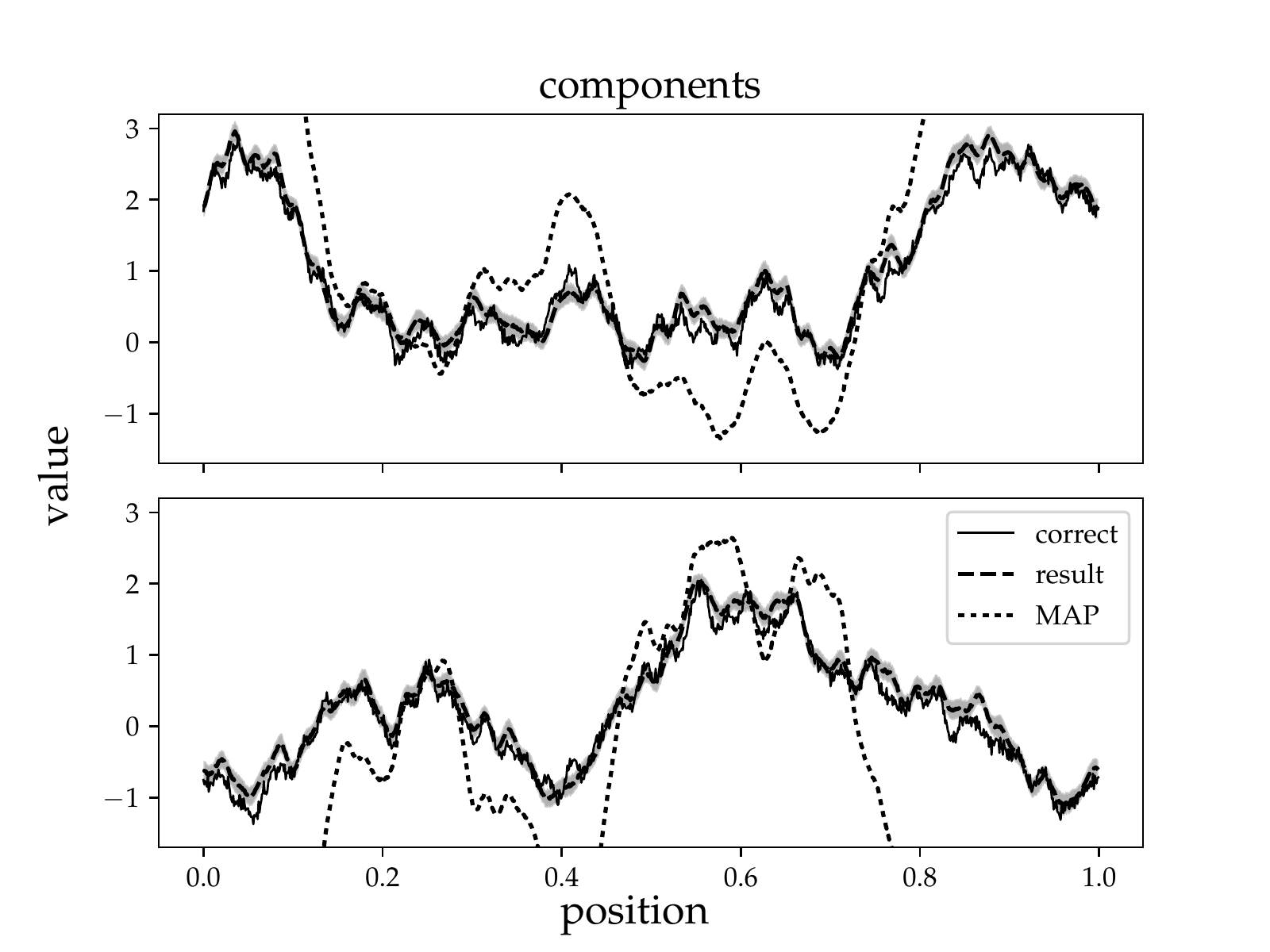}
\caption{Correct, reconstructed  and maximum posterior components with error estimate in scenario one of the data shown in Fig. 1.}
\label{fig:lowcomponents}
\includegraphics[scale = 0.56]{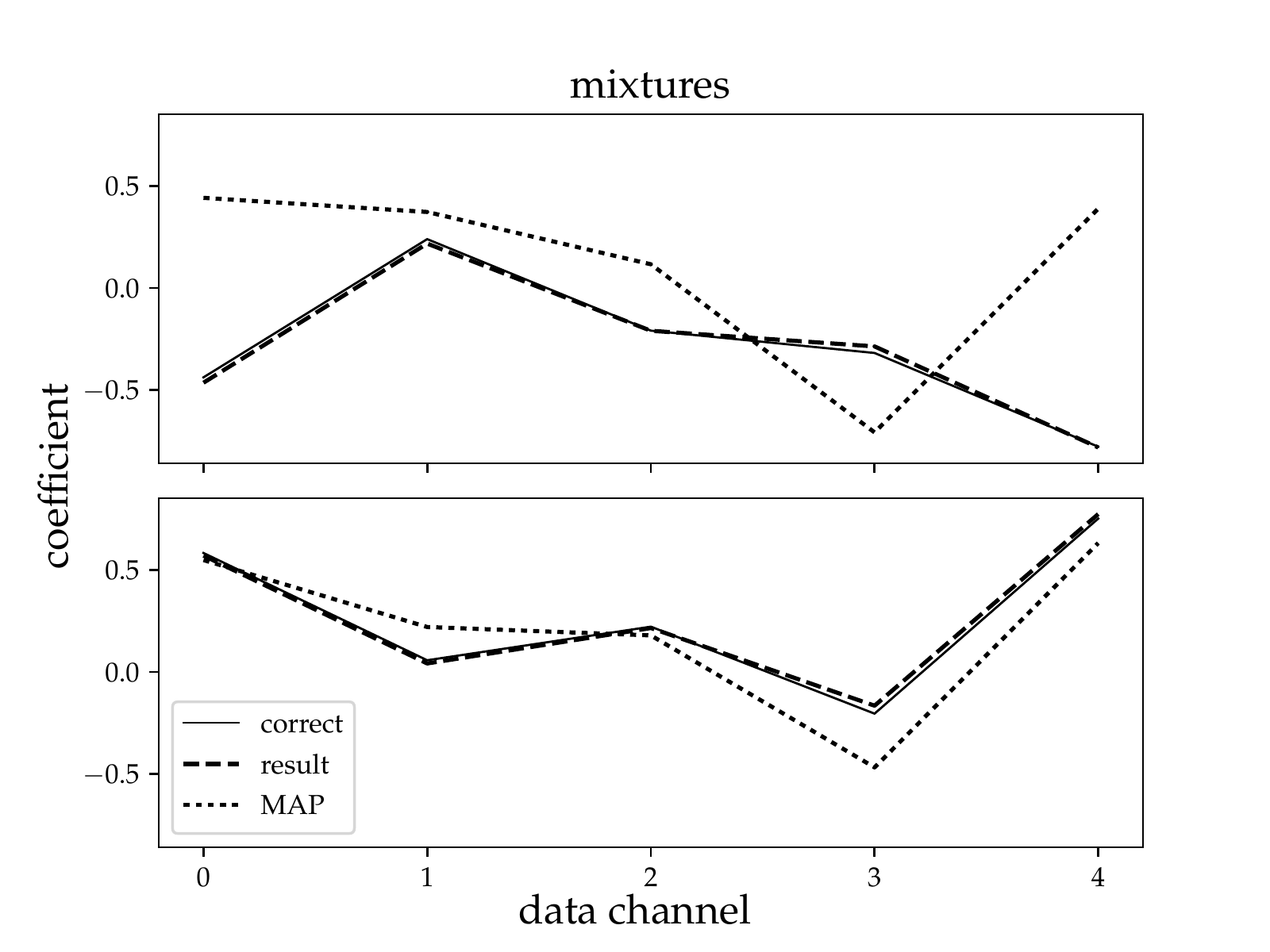}
\caption{Correct, reconstructed and maximum posterior mixtures in scenario one.}
\label{fig:lowmixture}
\end{center}

\end{figure}

\begin{figure}
\begin{center}
\includegraphics[scale = 0.56]{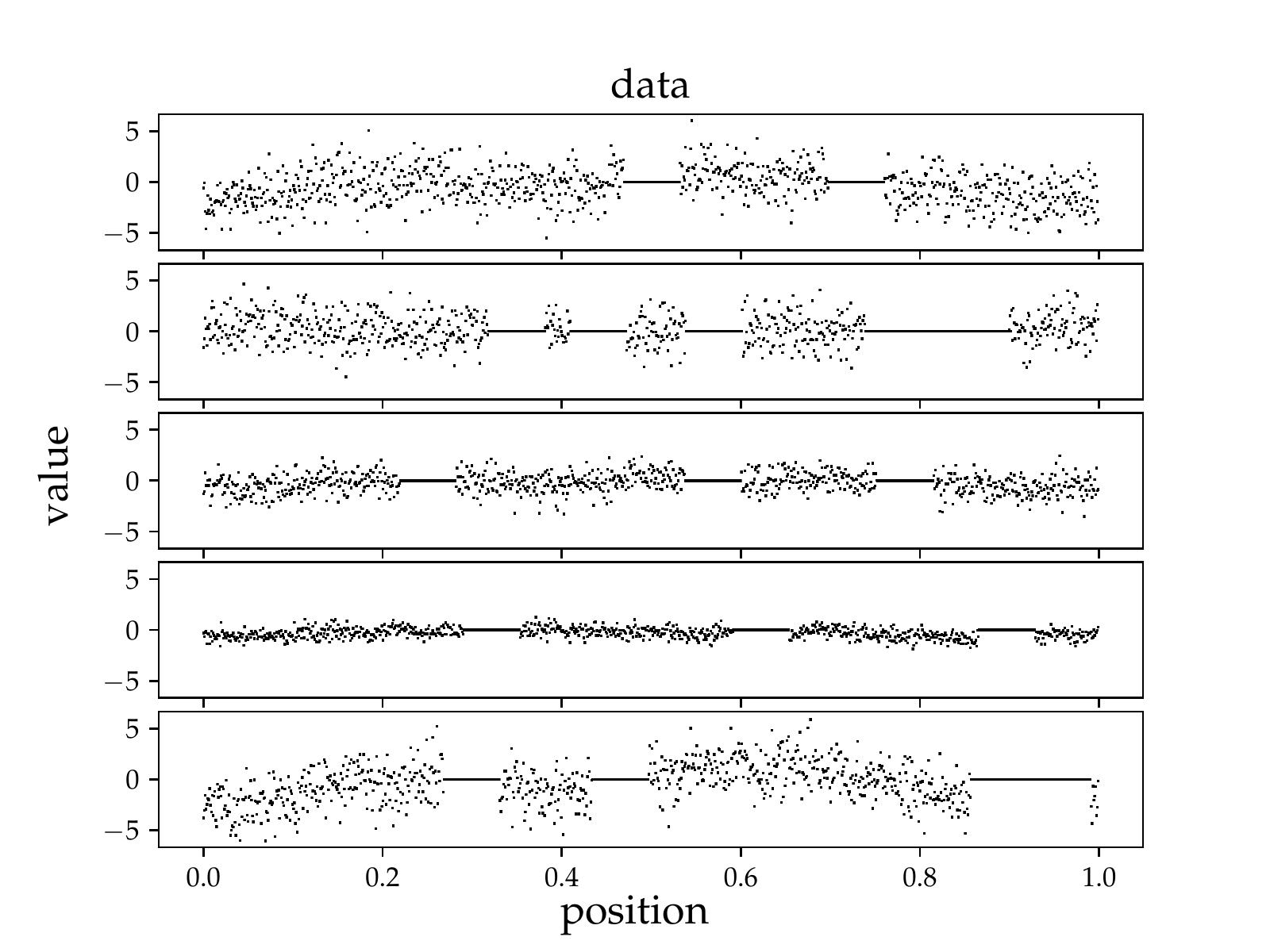}
\caption{Measurements with failing sensors and varying noise levels in scenario two. Note the changed scales.}
\label{fig:highdata}
\includegraphics[scale = 0.56]{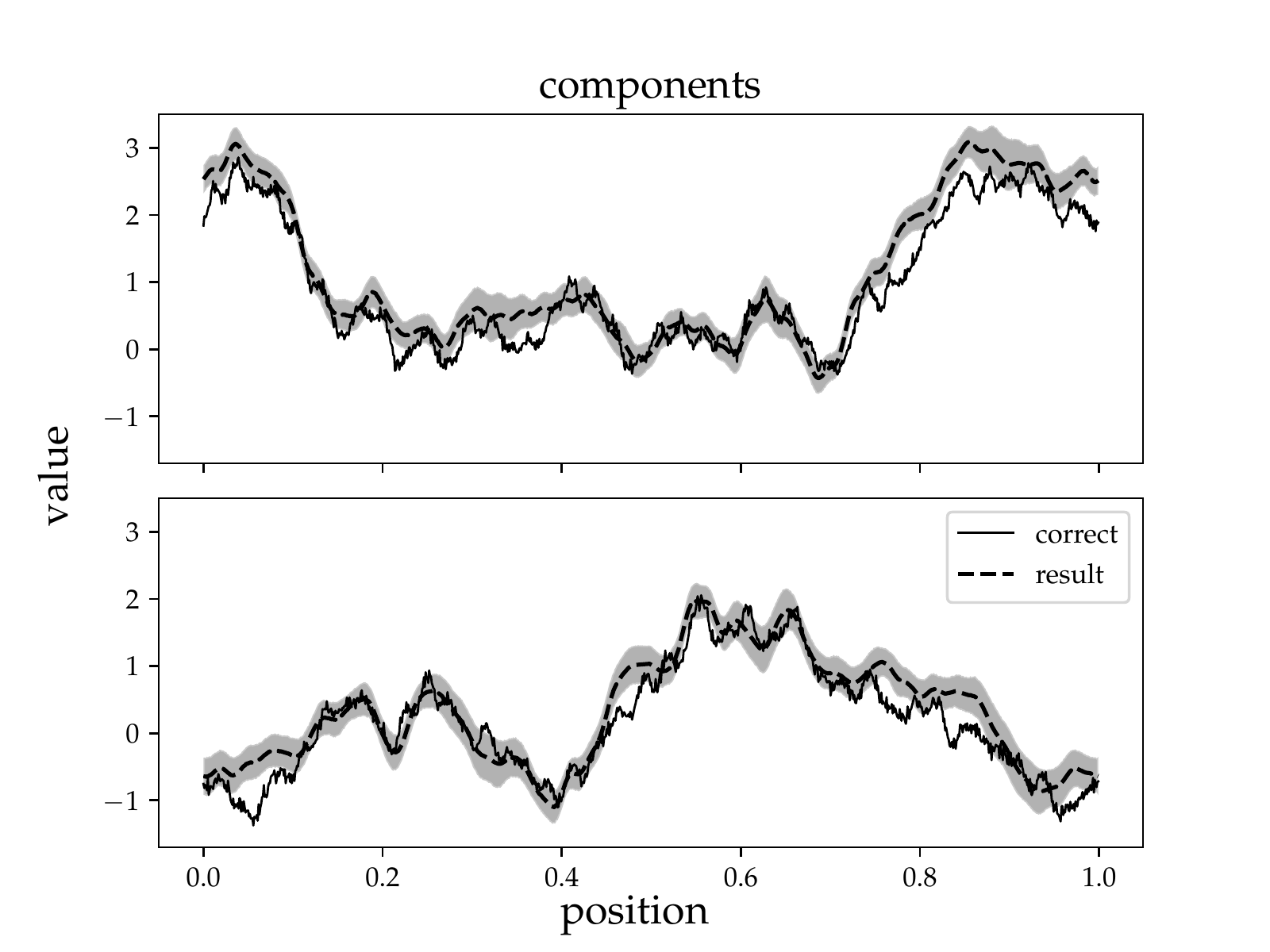}
\caption{Reconstruction of the independent components using the noisy data set of scenario two shown in Fig. 5. with error estimates. }
\label{fig:highcomponents}
\includegraphics[scale = 0.56]{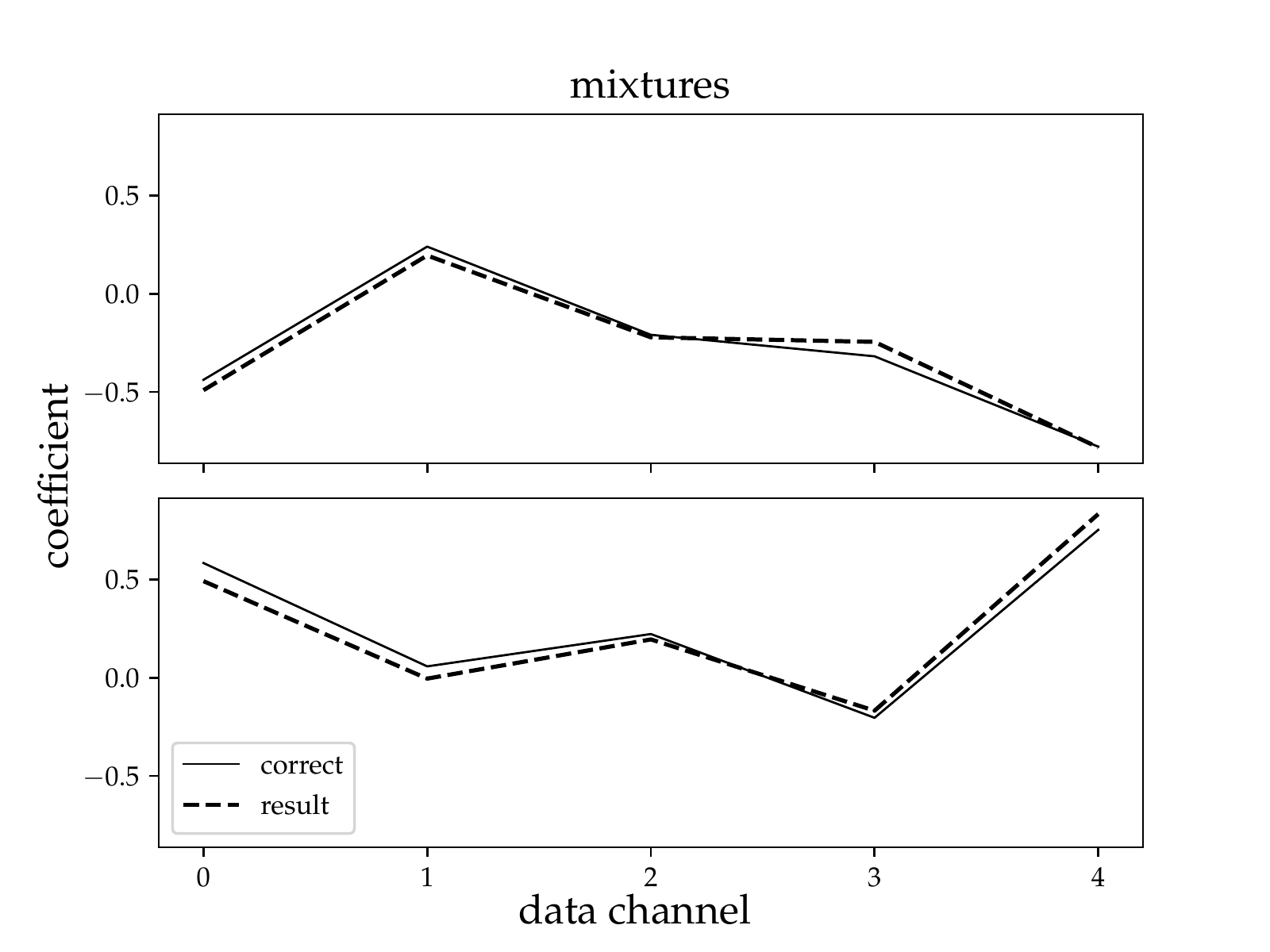}
\caption{Reconstructed mixtures from the noisy data set in scenario two.}
\label{fig:highmixture}
\end{center}

\end{figure}

We implemented the algorithm as outlined above in Python using the package \textsc{NIFTy} (Numerical Information Field Theory) \citep{Nifty}, allowing a coordinate free implementation. For our two numerical examples we will use synthetic data generated according to the model. The first one will describe a rather simple case with moderate, but present noise.

 In the second example we will challenge the algorithm with a more realistic measurement. We will model randomly failing measurement sensors by masking areas of the data set. In addition each sensor will exhibit a individual noise covariance of significantly increased strength.
 For the comparison we will use the same component realizations and mixture as used before.

  In our examples, we measure five different mixtures of two independent components. Each channel consists of $1024$ data points probing equally spaced locations of the unit interval over which our periodic components live. In the first example the measurements are corrupted by noticeable noise of zero mean and diagonal covariance of $\sigma_n^2=0.1$. The response operator $R$ in this case is just the identity operator $R_{xy} = \delta(x-y)$ . The data are illustrated in Figure \ref{fig:lowdata}. Both components are generated by drawing a realization from the prior distribution $\mathcal{P}(s)$ with power  spectrum 
\begin{align}
P_c(k) = \frac{1}{4k^2 + 1} \:\text{.}
\end{align}
This describes the spatial correlation by a falling power law in Fourier space\footnote{
We use here the numerical Fourier convention of $f(k) = \int_0^1 dx \:f(x) \:e^{2\pi ikx}$.
}
 which is typical for many physical processes.This function is shown in Figure \ref{fig:power}.
By choosing the same power spectrum for both components we can ignore the problem of multimodality of the probability distributions as all minima are equally global minima.
The values of the mixture entries are drawn independently from a Gaussian distribution with vanishing mean and unit variance. Afterwards the entries corresponding to one component are normalized to fix the multiplicative degeneracy between mixture and component.

The number of samples $s^*$  used to estimate the mixture was initially one per iteration and was increased to $25$ at the end of the reconstruction. We iterated the algorithm $300$ times, after which the reconstruction converged. The results of the analysis are shown in Figure \ref{fig:lowcomponents} and \ref{fig:lowmixture}. The reconstructions are corrected for the degeneracy of the signs and are compared with the true corresponding components and mixtures while keeping the product $Mm$ constant. We can clearly recover the morphological structure of the distinct components with high accuracy.  The one sigma uncertainty contours estimated as $\sqrt{D_{xx}}$ quantify the estimated error reasonably well.  The structure of the mixture is recovered, only small deviations from the true mixture are present. 
We can even recover relatively small  structures of the components as the algorithm uses the combined information of all channels simultaneously, increasing the effective signal-to-noise ratio, leading to higher resolutions. Denoising each individual channel first and then applying a noise free ICA method cannot reach that resolution as it is limited to the signal to noise ratio of the individual channels.

We also show the result of maximizing the posterior with respect to $s$ and $M$ for this scenario. The initial components are not recovered and the suggested solutions are highly anti-correlated. This demonstrates the necessity of the uncertainty corrections emerging from the presented model, represented by Eq. \ref{eq:corr} and the averages in Eq. \ref{eq:avrg}.
\\

In the second example we used the same setup as before with the same two components and five data channels. We only modified our measurement instrument to resemble typical properties of true sensors. We randomly masked $22\%$ of the total area by sequences of $64$ measurement points each.
Additionally we assign each sensor an individual noise covariance. The noise level will be significantly higher in this case, ranging from a factor of two up to $25$ times the variance compared to the previous example. The data are shown in Figure \ref{fig:highdata}. By eye it is hard to identify any components, only hints of correlated structures can be recognized.  We can encode the failing sensors in the instrument response operator $R$ as masks and the varying noise in the noise operator $N$ and run exactly the same algorithm as before. The result can be seen in Figure \ref{fig:highcomponents} and \ref{fig:highmixture}, again with corrected signs and compared to the true corresponding components. The morphological structure is recovered despite the significantly more hostile conditions. The overall uncertainty is consequently higher than before and therefore small scales are not as well resolved. Due to the masking we observe modulations in the uncertainty structure. In some parts the uncertainty does not fully cover the deviations from the true components. As we do not take the uncertainty  structure of the mixture into account we have probably underestimated the error. In the mixture we also observe larger deviations from the correct mixture, but in general we recover it well.
 
The convergence behavior for all three examples can be seen in Figure \ref{fig:convergence}. It shows the mean deviation of the current estimate $m_t$ of the components from the true components $s$ in each iteration step $t$, corrected for the degeneracy. We calculated it according to 
\begin{align}
\epsilon_t = \sqrt{\frac{(m_t - s)^\dagger(m_t-s)}{l}} \text{ ,}
\end{align}
where $l$ is the number of sites, given by the resolution of the components. We would expect this quantity not to become smaller than the expected deviations originating from the error estimate of the final result, which therefore sets the lower limit. It is shown as the two horizontal lines for the high and low noise case. During the inference the mean deviation declines towards this limit for both cases, but does not reach it.

This indicates that the error estimate of the result underestimates the error slightly, a finding that is not surprising as we do not take uncertainties of the mixture into account. The higher the noise level the more this effect becomes relevant, whereas in the low noise case it is almost negligible. We can also observe the statistical nature of the minimization due to the sampling in the noisy trajectory. Compared to that the maximum posterior minimization follows a smooth line. In this plot we can also nicely see the divergence of using just maximum posterior. It starts approaching the true components, roughly at the same speed as the KL-approach in the same situation, but then slows down and starts to accumulate errors and clearly diverges, while the other method continues converging. 
\section{Summary}
We derived a new method which allows for the separation of independent components from noisy measurements exploiting their auto-correlation. This was done by first describing the measurement process as a linear mixture of component fields which are observed by some linear measurement instruments under additive, Gaussian noise. From this model we derived the likelihood. Assuming homogeneity of the auto-correlation of the components we could express their correlation structure as a diagonal operator in the Fourier basis. From this assumption we derived the least informative prior distribution over the components in form of a Gaussian distribution. No prior assumptions about the mixture entries were made, but such could be added easily. Using the model likelihood and the component prior, we derived an expression for the posterior probability distribution by applying Bayes theorem. 

As this expression was not accessible analytically we approximated it by the product of a Gaussian distribution for the components and a delta distribution for the mixture entries. In order to infer the parameters of the approximation we proposed a scheme to minimize the Kullback-Leibler divergence of this distribution to the true posterior. It involved iterative Wiener filtering of the components and the mixture. For estimating the mixture we considered uncertainty corrections originating from the Gaussian approximation for the component maps. These turned out to be essential for obtaining accurate estimates of the mixture matrices. A joint MAP estimate of fields and mixtures tends to provide incorrect results. In order to evaluate the corrections we outlined an approach how to draw independent samples from the approximate Gaussian posterior distribution.

In two numerical examples we demonstrated the applicability of the derived algorithm. The first case involved moderate noise and recovered the true components and mixtures with high accuracy. The estimated error of the components was reliable. The second example models randomly failing sensors and  a significantly higher, varying noise level applied to the same components. The morphology of the mixture and components was recovered here as well, the error was slightly underestimated due to the involved point estimate of the mixture. Overall the algorithm delivered satisfying results and can also be applied in complex measurement situations in the high noise regime.
\section{Acknowledgments}
We acknowledge helpful discussions and comments on the manuscript by Martin Dupont, Reimar Leike, Sebastian Hutschenreuter, Natalia Porqueres, Daniel Pumpe, and two anonymous referees.

\bibliographystyle{apsrev4-1}

\bibliography{citations}

\end{document}